# Electronic ground state of Ni$_2^+$

V. Zamudio-Bayer, R. Lindblad, C. Bülow, G. Leistner, A. Terasaki, B. v. Issendorff, and J. T. Lau

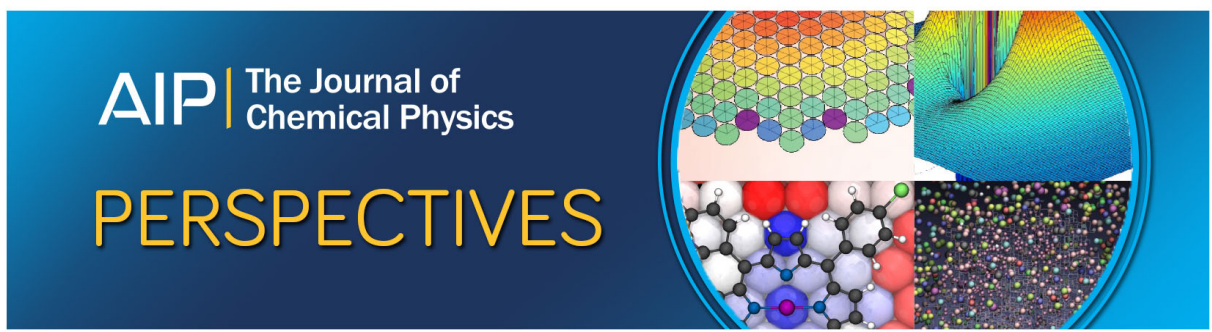

# Electronic ground state of $Ni_2^+$

V. Zamudio-Bayer,[1,2,a)] R. Lindblad,[1,3] C. Bülow,[1,4] G. Leistner,[1,4] A. Terasaki,[5,6] B. v. Issendorff,[2] and J. T. Lau[1,b)]

[1]*Institut für Methoden und Instrumentierung der Forschung mit Synchrotronstrahlung,*
*Helmholtz-Zentrum Berlin für Materialien und Energie, Albert-Einstein-Straße 15, 12489 Berlin, Germany*
[2]*Physikalisches Institut, Universität Freiburg, Stefan-Meier-Straße 21, 79104 Freiburg, Germany*
[3]*Synkrotronljusfysik, Lunds Universitet, Box 118, 22100 Lund, Sweden*
[4]*Institut für Optik und Atomare Physik, Technische Universität Berlin, Hardenbergstraße 36,*
*10623 Berlin, Germany*
[5]*Cluster Research Laboratory, Toyota Technological Institute, 717-86 Futamata, Ichikawa,*
*Chiba 272-0001, Japan*
[6]*Department of Chemistry, Kyushu University, 744 Motooka, Nishi-ku, Fukuoka 819-0395, Japan*

The $^4\Phi_{9/2}$ ground state of the $Ni_2^+$ diatomic molecular cation is determined experimentally from temperature and magnetic-field-dependent x-ray magnetic circular dichroism spectroscopy in a cryogenic ion trap, where an electronic and rotational temperature of $7.4 \pm 0.2$ K was reached by buffer gas cooling of the molecular ion. The contribution of the spin dipole operator to the x-ray magnetic circular dichroism spin sum rule amounts to $7 T_z = 0.17 \pm 0.06$ $\mu_B$ per atom, approximately 11% of the spin magnetic moment. We find that, in general, homonuclear diatomic molecular cations of $3d$ transition metals seem to adopt maximum spin magnetic moments in their electronic ground states. *Published by AIP Publishing.* [http://dx.doi.org/10.1063/1.4967821]

## I. INTRODUCTION

Homonuclear diatomic molecules and molecular ions have been studied since the first days of quantum mechanics. Surprisingly many of these, in particular those containing transition elements, are still far from being fully understood. Even though they pose a challenge to computational approaches in physics and chemistry, they are studied intensely because of a wide interest in their electronic, magnetic, and catalytic properties.[1–10] For example, the nickel diatomic molecular cation $Ni_2^+$ has served as a model system to study the role of surface defects in catalytic activity[11] and as one of the smallest systems to study strongly correlated electron phenomena.[12] Diatomic molecules of transition elements have also been proposed as systems with large magnetic anisotropy energy.[13–15] One prerequisite for large magnetic anisotropy energy is the large orbital angular momentum with significant spin-orbit coupling and strong coupling of the electronic orbital angular momentum to a molecular axis. The spin multiplicity $2S + 1$ and orbital angular momentum projection $\Lambda$ in the ground state of $Ni_2^+$ are, however, not known unambiguously.

A total angular momentum projection of $\Omega = 9/2$ onto the molecular axis and a bond length of $2.225 \pm 0.005$ Å to $2.242 \pm 0.001$ Å in the electronic ground state of $Ni_2^+$ were previously determined from rotationally resolved photodissociation spectroscopy.[16–19] Reported experimental values of the bond dissociation energy vary from $2.08 \pm 0.07$ eV to $2.32 \pm 0.02$ eV, with a preference for the upper values.[17,20,21] This experimentally determined $\Omega = 9/2$ ground state is in conflict with theoretical results, which predict $^4\Sigma$ and $^4\Delta$ states or quartet spin states without symmetry specification.[7,22–27] An excited $^4\Gamma$ state is also predicted at $\leq 30$ meV above a $^4\Sigma$ ground state.[22,23]

Here we show that the ground state of $Ni_2^+$ is the $^4\Phi_{9/2}$ state as determined from x-ray magnetic circular dichroism spectroscopy. The assignment agrees in the projected total angular momentum $\Omega = 9/2$ with photodissociation spectroscopy[16] and in the quartet spin state with theoretical predictions.[7,22–27]

## II. EXPERIMENTAL SETUP AND METHODS

The experimental setup has already been described in earlier reports of the electronic ground states of chromium,[28,29] manganese,[28,30] iron,[31] and cobalt[31] diatomic molecular cations. As in these previous studies, x-ray magnetic circular dichroism (XMCD) spectroscopy is performed at the Berlin synchrotron radiation facility BESSY II beamline UE52-PGM in a cryogenic linear ion trap that is situated in a $\mu_0 H = 5$ T homogeneous magnetic field of a superconducting solenoid.[29–36] In brief, $Ni_2^+$ ions are produced by magnetron sputtering of a nickel target with argon ions in helium buffer gas. Cationic species produced in the sputtering and gas-aggregation process are collected at the source exit and transferred by a differentially pumped radio-frequency hexapole ion guide into a quadrupole mass spectrometer.[37,38] A continuous and mass-filtered beam of $Ni_2^+$ is then guided into a liquid-helium-cooled quadrupole ion trap and thermalized in the presence of an axial magnetic field by collisions with helium buffer gas at constant pressure in the order of $10^{-4}$ mbar. Elliptically polarized soft x-ray radiation is coupled into the ion trap with the polarization vector parallel or antiparallel to the magnetic field, and the photon energy is swept across

a)Electronic mail: vicente.zamudio-bayer@physik.uni-freiburg.de
b)Electronic mail: tobias.lau@helmholtz-berlin.de

the nickel $L_{2,3}$ ($2p \rightarrow 3d$) absorption edges from 840.0 eV to 885.8 eV with a photon energy resolution of 400 meV in 0.2 eV steps. Auger decay following core excitation leads to multiply charged diatomic cations, which undergo fragmentation. Product ions eventually resulting from this resonant photoionisation process are collected by the ion trap and mass-analyzed by means of a reflectron time-of-flight mass spectrometer. X-ray absorption and x-ray magnetic circular dichroism spectra were recorded in a partial ion yield mode on the dominant $Ni^{2+}$ product ion channel. A total of nine data sets were taken under different conditions, yielding XMCD results at eight different ion temperatures and two different magnetic fields. The x-ray absorption and XMCD line shapes agree in all cases, only the XMCD intensity depends on magnetization.

### III. X-RAY MAGNETIC CIRCULAR DICHROISM SPECTROSCOPY OF $Ni_2^+$

The isotropic x-ray absorption spectrum of $Ni_2^+$, shown in Fig. 1, is taken as the sum of parallel ($I^+$) and antiparallel ($I^-$) orientation of photon helicity and applied magnetic field. It is in very good agreement with the previously reported linear x-ray absorption spectrum of $Ni_2^+$, except for the lower photon energy resolution and thus relative peak heights of the transitions.[39] As for the atomic $Ni^+$ cation and different from bulk, the direct $L_3$ photoionization threshold is significantly higher in energy than resonant $L_3$ transitions by $\approx$10 eV, leading to an absorption cross section that decreases almost to its pre-edge value between the absorption edges as can be seen around 860 eV excitation energy.[40] The x-ray absorption spectrum of Fig. 1 neither agrees with a pure $3d^8$ nor $3d^9$ ground state configuration[40] of the constituent atoms but seems to indicate a combination of both. This is illustrated by computed spectra for a linear combination of atomic $3d^8 4s^2$ and $3d^9$ initial state configurations in Fig. 1, calculated in a Hartree-Fock multiplet approach using Cowan's code as implemented in Missing.[41,42]

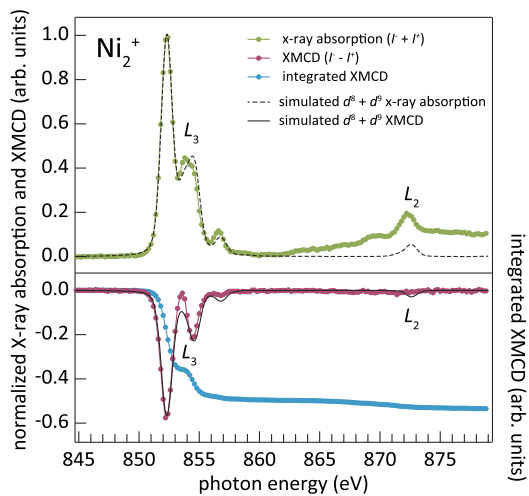

FIG. 1. $L_{2,3}$ edge X-ray absorption and XMCD spectra of $Ni_2^+$ recorded at $\mu_0 H = 5$ T applied magnetic field and $T = 3.8$ K ion trap temperature. The spectral profile of transitions into $3d$ derived final states can be reproduced by a combination of 0.35:1 Ni $3d^8 4s^2$ and $Ni^+$ $3d^9$ initial states. The smooth increase in the experimental x-ray absorption cross section above 862 eV is due to transitions into higher $nd$ ($n \geq 4$) and $ns$ ($n \geq 5$) derived Rydberg states.

Fig. 1 also shows the XMCD spectrum of $Ni_2^+$, i.e., the difference spectrum $I^- - I^+$, recorded at a magnetic field of $\mu_0 H = 5$ T and 3.8 K ion trap temperature. Because we did not observe a change of the XMCD line shape with magnetization of $Ni_2^+$ and because $I^-$ and $I^+$ are always positive values, the asymmetry at the $L_3$ edge varies linearly with magnetization and is limited to $|(I^- - I^+)/(I^- + I^+)| \leq 1$ in the case of saturation magnetization. It can therefore be used as a relative measure of magnetization, and the absolute value of the XMCD asymmetry at the maximum of the $L_3$ resonance ($\approx$ 852.25 eV) of $|(I^- - I^+)/(I^- + I^+)| = 0.58$ indicates a magnetization of $\geq 0.58$ times the total magnetic moment of $Ni_2^+$. At the same time, the large negative XMCD signal at the $L_3$ edge and the small but again negative XMCD signal at the $L_2$ edge, visible as a step in the integrated XMCD spectrum, indicate significant orbital magnetization and thus a large orbital contribution to the total magnetic moment. Application of the XMCD orbital and effective spin sum rules[43,44] to the spectrum in Fig. 1, recorded at lowest temperature of the ion trap and highest applied magnetic field, yields orbital and effective spin magnetizations, $m_\Lambda$ and $m_\Sigma$, per unoccupied $3d$-derived state of $m_\Lambda = 0.64 \pm 0.01~\mu_B$ and $m_\Sigma = 0.72 \pm 0.10~\mu_B$.

### IV. ELECTRONIC GROUND STATE OF $Ni_2^+$

#### A. Electronic configuration and candidate states

The lowest-energy asymptote of $Ni_2^+$ in the limit of separated atoms corresponds to Ni $3d^8 4s^2$ $^3F_4$ + $Ni^+$ $3d^9$ $^2D_{5/2}$ and leads to a $3d^{17} 4s^2$ electronic configuration of the diatomic molecular cation. Only 25 meV higher in energy is the Ni $3d^9 4s^1$ $^3D_3$ + $Ni^+$ $3d^9$ $^2D_{5/2}$ asymptote that would lead to a $3d^{18} 4s^1$ electronic configuration. Averaging over spin-orbit coupled states inverts the energetic order and places the latter asymptote 30 meV below the former.[20,45] We therefore consider both asymptotes in the following, even though the experimental x-ray absorption and XMCD spectra already indicate a contribution of atomic $3d^8$ and $3d^9$ configurations, making the $3d^{18} 4s^1$ configuration less likely the one that correctly describes $Ni_2^+$. Both asymptotes limit the spin multiplicity of $Ni_2^+$ to $2S + 1 \in \{2, 4\}$, and the projection $\Lambda$ of the orbital angular momentum onto the molecular axis is limited to $\Lambda \leq 5$ for the $3d^{17}$ configuration and $\Lambda \leq 4$ for the $3d^{18}$ molecular configuration.

#### B. Analysis of the orbital magnetization

XMCD sum rules[43,44] link the integrated intensities at the $L_3$ and $L_2$ edges of the XMCD asymmetry, depicted in the lower panel of Fig. 1, to expectation values $\langle L_z \rangle$ and $\langle S_z \rangle$ + $7/2 \langle T_z \rangle$ of the orbital and spin angular momenta, where $T_z$ is the intra-atomic spin dipole operator[46–50] that contributes to the XMCD effective spin sum rule. These expectation values are connected to magnetizations by $m_\Lambda = \mu_B/\hbar \langle L_z \rangle$ and $m_\Sigma + m_T = \mu_B/\hbar (2 \langle S_z \rangle + 7 \langle T_z \rangle)$. The orbital angular momentum sum rule[43] of XMCD is more robust than the effective spin sum rule[44] because the former does not suffer from $T_z$ contributions nor does it require a separation of $L_3$ and $L_2$ transitions in the experimental spectra.[51] Furthermore, the spin magnetic moment of $Ni_2^+$ might not be purely $3d$ derived but could have additional contributions from $4s\sigma$ orbitals, which $L_{2,3}$-edge XMCD is insensitive to. In contrast, only $3d$ derived $\pi$ and $\delta$ orbitals contribute to the orbital angular momentum of $Ni_2^+$.

As a first step, we therefore use the detected orbital magnetization $m_\Lambda$ as a measure of the hypothetical total magnetization of $Ni_2^+$ for each of the candidate electronic states. This is done by multiplying the highest experimental orbital magnetization obtained in the series of temperature and field-dependent XMCD spectra, $m_\Lambda = 0.64 \pm 0.01$ $\mu_B$ per unoccupied $3d$ state, with the number of unoccupied $3d$ states for the possible $3d^{17}\,4s^2$ ($m_\Lambda = 1.92 \pm 0.03$ $\mu_B$) and $3d^{18}\,4s^1$ ($m_\Lambda = 1.28 \pm 0.02$ $\mu_B$) configurations of $Ni_2^+$. These values are then normalized to the projected orbital magnetic moments $\Lambda$ of each candidate electronic state of the corresponding configurations. We compare the values of $m_\Lambda/\Lambda$ thus obtained for the different candidate states to the condition of $0.58 \leq m_\Lambda/\Lambda \leq 1$, where the lower limit of the relative magnetization is set by the XMCD asymmetry under the same experimental conditions, and the upper limit is set by saturation magnetization.

This analysis shows that $H$ ($\Lambda = 5$) and $\Gamma$ ($\Lambda = 4$) states of the $3d^{17}\,4s^2$ configuration as well as $\Gamma$ ($\Lambda = 4$) and $\Phi$ ($\Lambda = 3$) states of the $3d^{18}\,4s^1$ configuration are incompatible with the condition of a relative magnetization of $\geq 0.58$ derived from the XMCD asymmetry at the $L_3$ edge. These states can therefore be ruled out. Likewise, all $\Pi$ ($\Lambda = 1$) and $\Sigma$ ($\Lambda = 0$) states can be ruled out because the absolute value of the orbital magnetization cannot become larger than the orbital magnetic moment. Thus, without any further assumptions, the detected orbital magnetization and the limits of $1.92 \pm 0.03 \leq \Lambda \leq 3.31 \pm 0.05$ for the $3d^{17}\,4s^2$ configuration as well as $1.28 \pm 0.02 \leq \Lambda \leq 2.21 \pm 0.03$ for the $3d^{18}\,4s^1$ configuration only allow $\Delta$ ($\Lambda = 2$) and $\Phi$ ($\Lambda = 3$) states for the $3d^{17}\,4s^2$ configuration and $\Delta$ ($\Lambda = 2$) states for the $3d^{18}\,4s^1$ configuration.

### C. Ion temperature considerations

The ion temperatures that would correspond to these electronic states of the rotating $Ni_2^+$ molecular ion in the applied magnetic field are obtained by numerically solving an effective Zeeman Hamiltonian[52,53] at the respective magnetizations for Hund's case (a) angular momentum coupling.[31] We use the experimental value of $2.225 \pm 0.001$ Å of the equilibrium distance[16] to determine the rotational constant that is a necessary parameter. The ion temperature allows us to further exclude $\Delta$ ($\Lambda = 2$) states of the $3d^{17}\,4s^2$ configuration because these would correspond to ion temperatures of $\leq 1$ K, i.e., significantly lower than the lowest ion trap temperature of 3.8 K. At this stage, only $\Phi$ ($\Lambda = 3$) states of the $3d^{17}\,4s^2$ configuration and $\Delta$ ($\Lambda = 2$) states of the $3d^{18}\,4s^1$ configuration are left as potential candidates for the electronic ground state of $Ni_2^+$.

While the spin multiplicity could be 2 or 4 in both cases, only $3d$ derived electronic spins contribute to the multiplicity for the $3d^{17}\,4s^2$ configuration but also a $4s$ derived spin contributes in case of the $3d^{18}\,4s^1$ configuration. This single spin in the singly occupied $4s\sigma$ molecular orbital could be coupled parallel or antiparallel to the $3d$-derived spins, giving rise to $3d^{18}\,(^3\Delta)\,4s^1\,^2\Delta$ and $3d^{18}\,(^3\Delta)\,4s^1\,^4\Delta$ states. The $3d^{18}\,(^3\Delta)\,4s^1\,^2\Delta$ state with antiparallel spin coupling of the $3d$ and $4s$ derived states would correspond to an ion temperature of $3.4 \pm 0.4$ K, just below the ion trap temperature of 3.8 K, and can also be ruled out because of inevitable radio-frequency heating of the ions. Another possible state, $3d^{18}\,(^1\Delta)\,4s^1\,^2\Delta$, with an equal number of occupied $3d$ derived spin-up and spin-down states can also be ruled out because this state would lead to a vanishing $3d$ spin magnetization, in contradiction to the experimental finding of $m_\Sigma = 0.72 \pm 0.10$ $\mu_B$ per unoccupied $3d$ derived state. This reduces the remaining candidate states to $^2\Phi$ and $^4\Phi$ for the $3d^{17}\,4s^2$ configuration and $^4\Delta$ for the $3d^{18}\,4s^1$ configuration.

### D. Analysis of the ratio of orbital-to-spin magnetization

These remaining candidate states are assessed by the experimentally determined ratio of orbital-to-spin magnetization of $Ni_2^+$, which is equal to the ratio of orbital-to-spin magnetic moments in Hund's case (a) coupling of the angular momenta. This ratio of orbital to effective spin magnetic moment $\mu_\Lambda/\mu_\Sigma$ of $Ni_2^+$ is shown as a function of the orbital magnetization $m_\Lambda$ per $3d$ hole in Fig. 2, obtained by the sum rule analysis of a series of temperature and magnetic-field-dependent XMCD spectra of $Ni_2^+$. Except for the lowest magnetization, which was recorded at $\mu_0 H = 1.19$ T, these data were obtained at $\mu_0 H = 5$ T for different temperatures of the trapped ions by variations of buffer gas pressure, radio-frequency amplitude, and ion trap temperature. A weighted linear fit of $\mu_\Lambda/\mu_\Sigma$ yields an intercept of $0.86 \pm 0.12$ and a slope of $0.05 \pm 0.16$, i.e., zero slope within the error bars. Since a constant ratio is indeed expected in $LS$ coupling,[32,34] the ratio of $\mu_\Lambda/\mu_\Sigma$ can be given as $0.90 \pm 0.03$ from a weighted average that is shown as a dashed line in Fig. 2.

For the remaining states of the $3d^{17}\,4s^2$ configuration, values of $\mu_\Lambda/\mu_\Sigma = 3$ for the $^2\Phi$ state and $\mu_\Lambda/\mu_\Sigma = 1$ for the $^4\Phi$ state would be expected. For the $^4\Delta$ state of the $3d^{18}\,4s^1$ configuration, only the $^3\Delta$ contribution of the $3d$ derived subshell with a value of $\mu_\Lambda/\mu_\Sigma = 1$ can be compared with the experimental ratio to which the $4s$ derived spin does not contribute. As can be seen, none of the remaining candidate states has a value of $\mu_\Lambda/\mu_\Sigma$ that directly falls within the experimental range of $\mu_\Lambda/\mu_\Sigma = 0.90 \pm 0.03$. This indicates a non-negligible contribution of $T_z$ to the experimental ratio via the effective spin sum rule. These remaining states are therefore assessed by their $T_z$ contribution to the effective spin magnetic moment and by a radio-frequency heating contribution to the ion temperature.

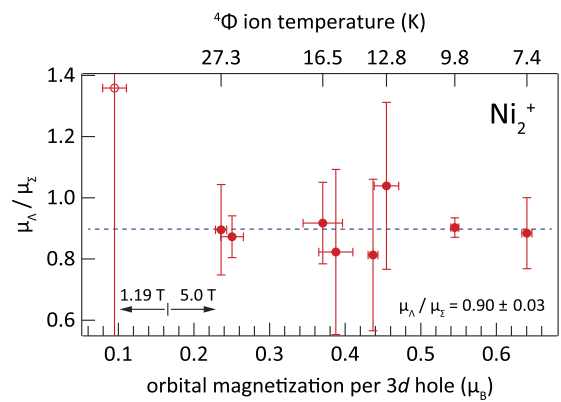

FIG. 2. Experimental ratio of $Ni_2^+$ orbital-to-spin magnetic moments as a function of the orbital magnetization, which was varied by ion temperature at constant $\mu_0 H = 5$ T. Only the value for the lowest orbital magnetization was obtained at $\mu_0 H = 1.19$ T. Ion temperatures that correspond to the $^4\Phi_{9/2}$ ground state are indicated in the upper axis for comparison.

The $^2\Phi$ state with more than three times the experimental $\mu_\Lambda/\mu_\Sigma$ value corresponds to an ion temperature of $4.9 \pm 0.1$ K, i.e., very low radio-frequency heating of only $1.1 \pm 0.1$ K, and would require a very large $T_z$ contribution of $7 T_z = 1.2 \pm 0.1$ $\mu_B$ per atom, more than twice its spin magnetic moment of 0.5 $\mu_B$ per atom. This state can therefore be ruled out. The $^4\Delta$ state would correspond to an ion temperature of $6.2 \pm 0.2$ K, and the $^3\Delta$ contribution of the $3d$-derived molecular orbitals, which XMCD is sensitive to, would require a small $T_z$ contribution of $7 T_z = -0.11 \pm 0.08$ $\mu_B$ per atom. However, the radio-frequency heating of only $2.4 \pm 0.2$ K is unexpectedly low at our experimental conditions.[29–32,34] Furthermore, the experimental spectrum in Fig. 1 clearly indicates a mixture of $3d^8$ and $3d^9$ initial state contributions that make a molecular $3d^{18} 4s^1$ configuration highly unlikely. This allows us to also rule out a $^4\Delta$ state. The remaining $^4\Phi$ state corresponds to an ion temperature of $7.4 \pm 0.2$ K, i.e., a reasonably low radio-frequency heating of $3.6 \pm 0.2$ K, and would require a small $T_z$ contribution to the effective spin of $7 T_z = 0.17 \pm 0.06$ $\mu_B$ per atom.

We therefore conclude that the $3d^{17} 4s^2 \; ^4\Phi$ state is the electronic ground state of $Ni_2^+$. This state is clearly in agreement with the $\Omega = 9/2$ ground state that was determined from rotationally resolved photodissociation spectroscopy.[16] It also agrees in the quartet spin state with available theoretical predictions.[7,22–27] The electronic and rotational temperature of $7.4 \pm 0.2$ K is among the lowest that have been reached by buffer-gas cooling of molecular ions.[54]

## V. DISCUSSION

Similar to the cases of $Cr_2^+$, $Mn_2^+$, $Co_2^+$, and possibly $Fe_2^+$,[29–31] the ground state of $Ni_2^+$ is characterized by the maximum spin multiplicity as a result of strong $3d$ exchange that favors the parallel alignment of unpaired electron spins. This seems to be a general property of homonuclear diatomic $3d$ transition metal cations. The contribution of $m_T = 0.11 \, m_\Sigma$ to the XMCD effective spin sum rule for $Ni_2^+$ is also in line with previous experimental values[31] of $m_T \approx 0$ $\mu_B$ for $Co_2^+$ and $|m_T| \leq 0.19 \, m_\Sigma$ for $Fe_2^+$, indicating generally small values of $m_T$ for homonuclear diatomic molecules of $3d$ transition elements.

A rather strong localization of the $3d$ derived orbitals in $Ni_2^+$ is deduced from the line shape of the x-ray absorption spectrum that can be reproduced by a combination of atomic $3d^8 4s^2$ and $3d^9$ initial state configurations. As was noted earlier, $Ni_2^+$ has the most complex $L_{2,3}$ edge x-ray absorption spectrum in the $Ni_n^+$ series and is significantly different from $Ni^+$ as well as from $Ni_3^+$.[39] This finding is consistent with only a small $3d$ orbital contribution to bonding in $Ni_2^+$ as was inferred from trends in bond length and bond dissociation energies along the $Ni_2^{(+,0,-)}$ series and was ascribed to the $3d$ orbital contraction.[19]

The ground state of neutral $Ni_2$ is not known unambiguously but theory hints at $^1\Sigma$ or mixed $^1\Sigma$ and $^3\Sigma$ states.[21,55,56] The $^4\Phi$ ground state of $Ni_2^+$ cannot be reached from any of these states in a one-electron transition. This is similar to the cases of $Cr_2$, $Mn_2$, and $Fe_2$[29–31] as well as to isolated vanadium, cobalt, and nickel atoms.

## VI. CONCLUSION

The electronic ground state of the $Ni_2^+$ diatomic molecular cation is determined as $^4\Phi_{9/2}$ from x-ray magnetic circular dichroism spectroscopy of cryogenic ions in a linear ion trap. This ground state of $Ni_2^+$ agrees in the total angular momentum projection $\Omega = 9/2$ with previous experimental results[16] and in the spin multiplicity $2S + 1 = 4$ with theoretical predictions.[7,22–27] $Ni_2^+$ is a further example of a homonuclear diatomic molecular cation of $3d$ transition elements that adopts the maximum spin multiplicity,[29,30] significant orbital angular momentum,[31] and small spin-dipole term[31] in its ground state.

## ACKNOWLEDGMENTS

Beam time for this project was granted at BESSY II beamline UE52-PGM, operated by Helmholtz-Zentrum Berlin. Technical assistance and support by Thomas Blume, Robert Schulz, Helmut Pfau, and François Talon are gratefully acknowledged. This project was partially funded by the German Federal Ministry of Education and Research (BMBF) through Grant No. BMBF-05K13Vf2. The superconducting solenoid was provided by Toyota Technological Institute. A.T. acknowledges financial support by Genesis Research Institute, Inc. R.L. acknowledges financial support from the Swedish Research Council through Grant No. 637-2014-6929. B.v.I. acknowledges travel support by Helmholtz-Zentrum Berlin.